\documentclass[3p,times,twocolumn]{elsarticle}

\def\pdfordvi{2} 
\def\ecrcpack#1{\ifcase #1 \or ecrc\or ecrc-mod\fi}

\usepackage{\ecrcpack{\pdfordvi}}

\volume{(to be published in ``Phi-to-Psi 2011'' Proceedings)~RUB-TPII-07/2011}
\firstpage{1}
\journalname{HEP-version}

\runauth{N.\, Stefanis et al.}

\jid{}

\jnltitlelogo{Nuclear Physics B Proceedings Supplement}

\usepackage{amssymb}
\def\colorpack#1{\ifcase #1 \or pdftex\or {}\fi}
\usepackage[\colorpack{\pdfordvi}]{color}
\definecolor{darkgreen}{rgb}{0,0.65,0}

\biboptions{sort,compress}

\usepackage[figuresright]{rotating}

\usepackage{amsfonts,amsmath,bm}

\def\pdfmappack#1{\ifcase #1 \or \pdfmapfile{+txfonts.map}\or {}\fi}
\pdfmappack{\pdfordvi}
\graphicspath{{./figs/}}
\def\pdforeps#1{\ifcase #1 \or pdf\or eps\fi}

\begin{document}
\begin{frontmatter}

\dochead{}


\title{Pion-Photon Transition Form Factor and Pion Distribution Amplitude in QCD:
       Facing the Enigmatic Behavior of the BaBar Data}

\author{N.~G.~Stefanis\corref{cor1}\fnref{jinr,rub}}
 \ead{stefanis@tp2.ruhr-uni-bochum.de}
  \cortext[cor1]{Corresponding author}
\author{A.~P.~Bakulev\corref{cor2}\fnref{jinr}}
  \cortext[cor2]{Speaker of the talk at the ``Psi-to-Phi'' Conference}
\author{S.~V.~Mikhailov\fnref{jinr}}
\author{A.~V.~Pimikov\fnref{jinr}}
\address[jinr]{Bogoliubov Laboratory of Theoretical
Physics, JINR, 141980 Dubna, Russia}
\address[rub]{Institut f\"{u}r Theoretische Physik II,
              Ruhr-Universit\"{a}t Bochum, D-44780 Bochum, Germany}
\begin{abstract}
We present an extended analysis of the data
for the pion-photon transition form factor
from different experiments, CELLO, CLEO, and BaBar, and
discuss various theoretical approaches which try to reason from them.
We focus on the divergent behavior of the BaBar data for the pion
and those for the $\eta(\eta')$ pseudoscalar mesons and comment on
recently proposed explanations for this discrepancy.
We argue that it is not possible at present to accommodate these data
within the standard QCD framework self-consistently.
\end{abstract}


\end{frontmatter}



\section{Two-photon processes for $\pi^0$ and $\eta(\eta')$ in QCD}
\label{sec:intro}
At the basis of applications of Quantum Chromodynamics (QCD),
which governs the interactions of quarks and gluons,
is the property of factorization of corresponding amplitudes in hard
processes.
Proving the property of factorization means that a hadronic process
at large momentum transfer can be dissected in a product of distinct
quantities each associated with separate regimes of dynamics.
Then, subprocesses developing at short distances and involving
partons---quarks and gluons---can be accurately described in a
systematic way within perturbation theory.
On the other hand, the dynamical (mainly nonperturbative)
features of the hadron binding effects are encoded in correlators,
which contain quark and gluon field operators, and can be parameterized
in terms of light-cone wave functions and parton distribution
functions.
These quantities are process-independent and describe the universal
interpolation between hadrons and their quark and gluon degrees of
freedom as distributions over the fractions of the longitudinal and
intrinsic transverse momenta carried by the partons.
They have to be determined by nonperturbative methods (models),
lattice calculations, or from the data.

The (spacelike) transition form factors (TFFs) of pseudoscalar mesons,
in particular the pion, have been extensively studied within QCD,
because in leading order they are purely electromagnetic processes
with the binding QCD effects being factorized out into the pion
distribution amplitude (DA)---see \cite{BL89} for a review and
\cite{BMPS11} for recent references.
This means that for a highly virtual photon with the four-momentum
transfer $Q^2$ and a quasi-real photon with $q^2\to 0$, the TFF
can be cast as the convolution of the twist-two
hard-scattering amplitude
$T(Q^2,q^2\to 0,x)=Q^{-2}(1/x+O(\alpha_s))$,
describing the elementary process
$\gamma^*\gamma\to q\bar{q}$,
with the twist-two pion DA $\varphi^{(2)}_{\pi}(x;\mu^2)$
so that
\begin{eqnarray}
 \!\!\!\!\!\!\!\!\!
 Q^2 F^{\gamma^{}\gamma^{*}\pi}(Q^2)
  \!\!&\!=\!&\!\! \frac{\sqrt{2}}{3}f_\pi\!
     \int\limits_{0}^{1}\!\!Q^2 T(Q^2,x)\,
      \varphi^{(\text{2})}_{\pi}(x;Q^2)\,dx
\nonumber \\
  \!\!&\!\!&\!\!
   +\,O\left(\frac{\delta^2}{Q^2}\right)\,,
\label{eq:convolution}
\end{eqnarray}
where $\delta^2$ is the scale of the twist-four term
taking values in the range
$\delta^2 = 0.15 \div 0.23$~GeV$^2$.
Note that we have omitted for simplicity
variables irrelevant for our discussion
(see \cite{LB80} and \cite{MS09} for more details).

Several experimental groups have measured
$Q^{2}F^{\gamma^*\gamma\pi^0}(Q^2,q^2\to 0)$ and
$Q^2F^{\gamma^*\gamma\eta(\eta')}(Q^2,q^2\to 0)$
in two-photon processes
$e^+e^- \to e^+e^- \gamma^*\gamma \to e^+e^- X$,
where $X=\pi^{0}$ \cite{CELLO91,CLEO98,BaBar09},
$\eta$ and $\eta'$ \cite{CLEO98,BaBar11-BMS}.
The range of probed photon momentum varies from
$0.7 \div 2.2$~GeV$^2$ (CELLO),
to $1.64 \div 7.90$~GeV$^2$ (CLEO), to
$4.24 \div 34.36$~GeV$^2$ (BaBar).
A recent compilation and discussion of the experimental data
with the focus on the BaBar data can be found in \cite{Muller11}.
The current situation from the experimental point of view can be
summarized like this: all data for the TFFs pertaining to the
non-strange part of the $\eta$ and $\eta'$ conform with the asymptotic
QCD limit \cite{LB80}
$\lim_{Q^2\to\infty} Q^2F^{\gamma^*\gamma\pi^0}(Q^2)
 \longrightarrow
 \sqrt{2}f_\pi
 = 0.185
$
and are best described by endpoint-suppressed DAs,
while there is remarkable contrast to the high-$Q^2$ tail of
the BaBar data for the $\pi^0$ TFF \cite{BaBar11-BMS,Muller11}.
The rising trend of the BaBar data can be reproduced by the
fit \cite{BaBar09}
$Q^2F(Q^2)= A(Q^{2}/10~\text{GeV}^{2})^\beta$
with
$A=0.182 \pm 0.002$~GeV
and
$\beta = 0.25 \pm 0.02$.
The latter value differs significantly from 0
predicted by QCD perturbation theory,
cf. Eq.\,(\ref{eq:convolution}).

From the theoretical side,
the conclusions drawn from different approaches
are at odds with each other
and cannot accommodate all available data simultaneously
to resolve the puzzle.
In our recent data analysis in \cite{BMPS11}
(see also \cite{BMPS11strba}),
we used Light-Cone Sum Rules (LCSRs) \cite{Kho99,BBK89}
and performed an extended calculation of the pion TFF
which takes into account the next-to-leading (NLO) order
radiative corrections \cite{DaCh81}
and the twist-four contribution \cite{Kho99,BF89},
while also including in terms of uncertainties the main
next-to-next-to-leading-order (NNLO) correction \cite{MMP02,MS09}
and the twist-six term, computed in \cite{ABOP10}.
The evolution of $\varphi^{(2)}_{\pi}(x;\mu^2)$
with $\mu^2>1$~GeV$^2$
was also taken into account at the NLO level
with $\Lambda_\text{QCD}^{(3)}=370$~MeV
and
$\Lambda_\text{QCD}^{(4)}=304$~MeV.
The main observations are listed below.

(i) All data for $Q^2F^{\gamma^*\gamma\pi^0}(Q^2)$ in the range
$[1 \div 9]$~GeV$^2$
can be described at the level of $\chi^2 < 1$
with only two Gegenbauer coefficients
$a_2$ and $a_4$
with negative $a_4$,
satisfying $|a_4|\lesssim a_2$.

(ii) The error ellipses of the data (CELLO, CLEO, BaBar) in the
$(a_2,a_4)$ plane overlap with the allowed region for these two
parameters determined before \cite{BMS01} using QCD sum rules with
nonlocal condensates \cite{MR89}.
These pion DAs are characterized by a strong suppression of their
kinematical endpoints $x=0,1$, where $x$ is the longitudinal momentum
fraction of the quark inside the pion.

\begin{figure}[h!]
 \centerline{\hspace{0mm}\includegraphics[width=0.48\textwidth]{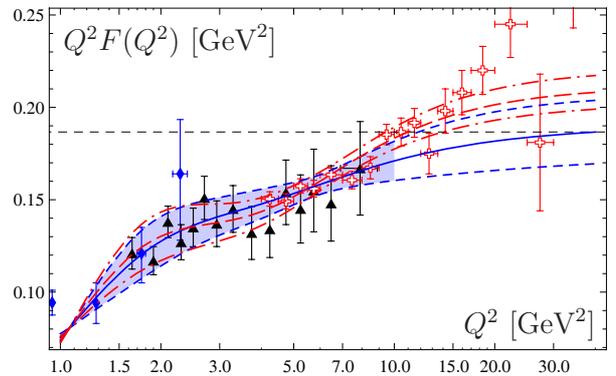}}
 \caption{\label{fig.pionFF3-D}
 Best-fit curves to the experimental data for the TFF in
 the framework of LCSRs, inside corresponding bands of
 68\% CL regions as sums of statistical errors and
 twist-four uncertainties.
 Blue lines refer to set~1 and red lines to set~2 of the data from
 \protect\cite{CELLO91,CLEO98,BaBar09},
 with designations as in Fig.\ \protect\ref{fig.pionFF.fit}.
 The horizontal dashed line marks the asymptotic QCD limit.}
\end{figure}

(iii) Beyond 10~GeV$^2$, an acceptable statistical description
of the data ($\chi^2 \geqslant 1$)
demands the inclusion of
a sizeable and positive Gegenbauer coefficient $a_6$
with $a_6\simeq-1.8\,a_4\simeq 1.7\,a_2$ (at the scale
$\mu_{\rm SY}=2.4$~GeV \cite{SY99}).
Still higher coefficients have been taken into account in
\cite{BMPS11strba} but found to have little effect.
Moreover, it is difficult to select the optimal number of harmonics
needed.
Thus, the endpoint enhancement provided this way is not
sufficient to reproduce the steep rise of the pion TFF observed by
BaBar \cite{BaBar09}.
The range of variation of the associated pion DAs,
conforming with this 3D ($a_2, a_4, a_6$) analysis,
was worked out for both sets of data, i.e.,
$[1 \div 9]$~GeV$^2$ --- set~1 and $[1 \div 40]$~GeV$^2$
--- set~2,
in \cite{BMPS11strba}, see also Fig.\ \ref{fig.pionFF3-D}.
This figure shows best-fit results for both data sets
in the form of bands of TFFs with errors stemming from the
sum of the statistical errors and the twist-four
uncertainties.
The best-fit curve to set~1 in Fig.\ \ref{fig.pionFF3-D}
is represented by the solid (blue) line, whereas the best-fit
curve to set~2 at high $Q^2$ is denoted by the dashed (red) line.
This behavior makes it apparent that in the framework of
LCSRs the fit to the BaBar data above 9~GeV$^2$
deviates from that at low-$Q^2$ at the level of 1$\sigma$.

Moreover, following \cite{Ste08} with some refinements explained in
\cite{BMPS11}, one obtains the following windows for the
values of the moments:
$\langle \xi^2 \rangle_\pi \in [0.23 \div 0.29]$,
$\langle \xi^4 \rangle_\pi \in [0.102 \div 0.122]$
and
$\langle \xi^2 \rangle_\pi \in [0.26 \div 0.29]$,
$\langle \xi^4 \rangle_\pi \in [0.11 \div 0.122]$.
These estimates were derived from our LCSR analysis
\cite{BMPS11,BMPS11strba} by combining all data of set~1
with the lattice results obtained in \cite{Lat06} and \cite{Lat10},
respectively.
All values were evaluated at the typical lattice scale
$\mu^2_\text{Lat}=4$~GeV$^2$,
using the definition
$\langle \xi^{N} \rangle_{\pi}\equiv
 \int_{0}^{1}\!\varphi_{\pi}^{(2)}(x,\mu^2_\text{Lat})\,(2x-1)^{N}dx
$,
with $\varphi_{\pi}^{(2)}(x, \mu^2)$ being normalized to unity.
The ``window'' for $\langle \xi^2 \rangle_\pi$, obtained for the
data with $Q^2<10$ GeV$^2$ (set~1),
has in terms of the $1\sigma$ error ellipse a large intersection
with the most recent lattice estimates
from \cite{Lat06,Lat07,Lat10}.
Including into the fit the high-$Q^2$ BaBar data, this intersection
significantly deteriorates---see \cite{BMPS11,BMPS11strba} for further
details.

(iv) These findings indicate a significant discrepancy between the
BaBar data for $\pi^0$ and the method of LCSRs---and in more general
terms the QCD factorization---at high-$Q^2$ values, an indication
that the analysis in \cite{ABOP10} is possibly no turning
point in understanding the high-$Q^2$ $\pi^0$ BaBar data within QCD.
Thus, our analysis does not confirm the opposite conclusions
drawn in \cite{ABOP10} which uses the same calculational scheme,
but a larger value of the auxiliary Borel parameter $M^2$,
notably 1.5~GeV$^2$, instead of values $M^2<1$~GeV$^2$
as in our approach, (which follows \cite{Kho99})
and more coefficients in the conformal expansion of the pion DA.

\begin{figure}[h!]
 \centerline{\hspace{0mm}\includegraphics[width=0.48\textwidth]{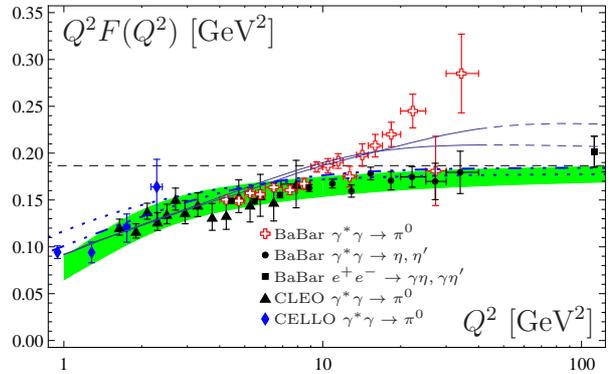}}
  \caption{\label{fig.pionFF.fit}
    Theoretical predictions for the scaled transition form factors for
    $\pi^0$ and the non-strange part of the $\eta$ and $\eta'$ from various
    theoretical approaches.
    The green strip contains the results obtained in
    \protect\cite{BMPS11} using the method described in the text.
    The two solid (blue) lines represent the findings of
    \protect\cite{ABOP10}, whereas the dotted and the
    double-dotted-dashed lines denote two independent predictions
    derived from AdS/QCD in \protect\cite{BCT11b} and
    \protect\cite{GR08}, respectively.
    The experimental data are taken from various experiments,
    referenced in the text.
         }
\end{figure}

(v) Conformity with the increasing trend of the BaBar data
can be actually achieved only with a flatlike pion DA,
like that proposed in \cite{Rad09}
and in a different context in \cite{Pol09}.
However, the use of such a pion DA
reduces the accuracy of the predicted TFF
at lower values of $Q^2$ \cite{BMPS11,MS09Trento}.
Moreover, it was emphasized in \cite{BMPS11,BMPS11strba,SteLC2011}
that then one becomes unable
to reproduce the BaBar data \cite{BaBar11-BMS}
for the nonstrange part of the $\eta$:
$|n\rangle=\left(1/\sqrt{2}\right)
 \left(|u\bar{u}\rangle + |d\bar{d}\rangle\right)$.\footnote{
 The separation of the nonstrange part of the $\eta$ and
 the $\eta'$ is model dependent.
 Moreover, the decay parameters $f_{\eta}$ and $f_{\eta'}$
 are not well known and strongly depend on the mixing
 between the two eta mesons.
 We here employed the mixing scheme of Ref.\ \cite{FKS98}.}
Indeed, employing the mixing scheme of \cite{FKS98},
one can relate the TFF of $|n\rangle$ to that of the $\pi^0$
multiplied by a factor $5/3$ due to the quark charges
(also assuming that $f_n=f_\pi$).
Hence, it appears that the TFF for the $\pi^0$ and the $|n\rangle$
follow antithetic trends that correspond to DAs with distinct
endpoint characteristics: extreme endpoint enhancement for the first
vs. endpoint suppression for the second \cite{SteLC2011}.
Because the properties of $|n\rangle$ are not so sensitive
to the choice of the mixing angle, such a strong antithetic behavior
cannot be ascribed to this uncertainty.
These results are shown in Fig.\ \ref{fig.pionFF.fit} using a
logarithmic scale for $Q^2$ and comparing them with all the available
experimental data.
Our predictions \cite{BMPS11} are represented by a (green) strip
whose width is a measure of the involved theoretical uncertainties,
while the solid (blue) lines reproduce and extend farther out
the predictions of \cite{ABOP10}.
The other lines will be explained shortly in Sec.\ \ref{sec:expl}.
The crucial question is:
What kind of mechanism underlies such a strong flavor-symmetry breaking
in the pseudoscalar meson sector of QCD?

In the next section, we will discuss some proposed explanations of
the BaBar data without attempting to be comprehensive or conclusive.

\section{Data Explanations}
\label{sec:expl}
Let us start by recalling some facts related to the pion TFF.
The CLEO data favor a pion DA close to the asymptotic (as) form
\cite{KR96,SSK99} excluding pion DAs of the Chernyak--Zhitnitsky (CZ)
\cite{CZ84} type.
Later, more detailed analyses of these data
\cite{SY99,BMS020305} have excluded $\varphi_{\pi}^\text{CZ}$
and $\varphi_{\pi}^\text{as}$ at the $4\sigma$ and the $3\sigma$ level,
respectively.
To achieve an agreement with the CLEO data with $1\sigma$ accuracy,
one has to use endpoint-suppressed pion DAs of the form derived in
\cite{BMS01} with the help of QCD sum rules with nonlocal condensates.
In view of these findings, it was thought that more precise
measurements which would extend the range of $Q^2$ to much higher
values, would confirm this trend because for large momentum transfers
the application of perturbative QCD on the basis of collinear
factorization should work better and better.
Surprisingly, the rapid growth of the high-$Q^2$ BaBar data is at odds
with these expectations, though this was not immediately recognized.
But in coincidence with the publication of the BaBar data in the arXiv,
two of us have clearly stated that no agreement between the standard
QCD scheme and a rising TFF (scaled by $Q^2$) can be achieved in terms
of endpoint-vanishing $\pi$ DAs \cite{MS09}.

Interestingly, other authors proposed at the same time
explanations of such an ``anomalous'' behavior
by appealing to flatlike pion DAs
\cite{Rad09,Pol09,Dor11},
outside the standard QCD context.
Such approaches suffer from the point of view
that they depend in a sensitive way
on specific, i.e., \textit{contextual} nonperturbative scales
that cannot be linked to the standard QCD scales.
Therefore, the meaning of these scales,
with their particular values
needed to explain the BaBar data,
remains obscure.
Consequently, though the BaBar data can be explained
within such schemes,
because they provide logarithmic enhancement to the TFF,
it is not possible
to identify a \textit{single} underlying physical mechanism
which yields enhancement of the pion TFF.
Moreover, it is difficult to understand
why the pion and the $\eta$ should behave
like ``pointlike'' particles
(see, for instance, \cite{RRBGGT10})
as implied by flatlike (or flat-top)
DAs,
while the DAs of the $\eta'$ and the $\eta_\text{c}$
should be close to their asymptotic forms
(eventually with additional endpoint suppression).\footnote{
Note parenthetically that the TFF for the $\eta_\text{c}$ approaches
the QCD prediction from below---see \cite{Kro11} for a discussion.}
The assertion in \cite{Dor11} that this is due to the larger mass of
these particles is not very convincing, given that the DAs of the
$\pi^0$ and the $\eta$ are very similar in shape in the nonlocal
chiral quark model of \cite{Dor11}, though the corresponding masses
$m_{\pi}=140$~MeV and $m_{\eta}=548$~MeV
are very different.
Nevertheless, it would be premature to dismiss the correctness and/or
relevance of such explanations.

The calculation of the $\pi^0$, $\eta$, and $\eta'$ TFFs
has been carried out within holographic approaches of AdS/QCD,
e.g., \cite{BCT11b}, \cite{GR08}, and \cite{SZ11}.
In Fig.\ \ref{fig.pionFF.fit} the two broken (blue) lines
represent the independent findings
of two such approaches for
$Q^2F^{\gamma^*\gamma\pi^0}(Q^2)$.
The dotted line denotes the prediction derived in \cite{BCT11b},
using a dressed electromagnetic current and taking into account
the twist-two and the twist-four hadronic AdS components
of the pion wave function (Eq.\ (43) in \cite{BCT11b})
within a soft-wall holographic approach.
The combination of the nonperturbative bound-state dynamics,
predicted by the holographic AdS/QCD correspondence,
with the perturbative Efremov--Radyushkin--Brodsky--Lepage evolution~\cite{LB80,ER80}
was treated in detail in \cite{BCT11}.
The double-dotted-dashed line shows an analogous result \cite{GR08},
obtained by an extension of the hard-wall AdS/QCD model which includes
the Chern--Simons term, required to reproduce the chiral anomaly of
QCD.
Note that the shown curve beyond 10~GeV$^2$ was generated by us.
It is obvious that both displayed predictions are incongruent
with the BaBar data for $\pi^0$, while they agree with each other and
with the BaBar data for the $\eta(\eta')$, though they somewhat
overestimate all data at lower values of $Q^2$.
Moreover, one observes that both predictions overlap with the band of
the results derived in \cite{BMPS11} and indicate conformity with
endpoint-vanishing (or even endpoint-suppressed) pion DAs,
like in the BMS formalism \cite{BMS01,BMS020305}.

The chiral anomaly plays an important role also in another approach,
proposed in \cite{KOT11b},
which combines an exact nonperturbative sum rule,
following from the dispersive representation of the axial anomaly,
and quark-hadron duality.
Within this approach it is claimed \cite{KOT10plb}
that the increase of the BaBar data for the $\pi^0$ is due to
small corrections to the continuum
(i.\,e., an infinite number of higher resonances)
which entail a strong enhancement of the pion TFF, whereas the same
effect for the $\eta$ turns out to be several times
smaller.\footnote{
 Note that until now, all calculated corrections,
 perturbative and nonperturbative, amount in total to a negative
 contribution so that it is not clear how this scenario can be
 realized.}
Depending on the particular mixing scheme adopted, the obtained
predictions \cite{KOT11b} for the $\eta$ and $\eta'$ mesons
agree with the gross of the BaBar data for the associated TFFs.
The usefulness of the dispersive approach \cite{KOT10plb}
---which does not rely upon factorization---has to be further tested
by extending it to the singlet channel.
Another approach that relates the BaBar effect to the chiral
anomaly is discussed in \cite{PP11}.

The light pseudoscalar meson-photon TFFs for the pion and the $\eta$
and $\eta'$ mesons were also discussed in \cite{WH11}, using a
quark-flavor mixing scheme with only one mixing angle \cite{Fel00}.
The associated DAs of these mesons are obtained from a light-cone wave
function \cite{WH10} by tuning a master parameter to appropriate
values, whereas the constituent quark masses and the mixing angle
play a relatively minor role.
The main message from this analysis is that the (logarithmic) growth of
the scaled TFF $Q^2F^{\gamma^*\gamma\pi^0}(Q^2)$, indicated by the
BaBar data, cannot be reproduced, while it remains unexplained why
$Q^2F^{\gamma^*\gamma\pi^0}(Q^2)$ and
$Q^2F^{\gamma^*\gamma\eta}(Q^2)$
should behave so differently.
In fact, within the range of the model parameters, the BaBar data on
$Q^2F^{\gamma^*\gamma\eta}(Q^2)$ and $Q^2F^{\gamma^*\gamma\eta'}(Q^2)$
can be explained simultaneously in the whole $Q^2$ region.
So far, no single mechanism with a physical interpretation within the
standard QCD framework has been identified to explain the increase of
the pion TFF as a result of particular quark-gluon interactions.
Effects associated with the transverse-momentum degrees of freedom,
intrinsic, i.e., inside the pion wave function, and resummed in terms
of Sudakov factors, have also been discussed with respect to the BaBar
data, see, e.g., \cite{LiMi09,BBKI09,Kro11}.
Some other examples of studies of the BaBar data in conjunction
with particular nonperturbative QCD models
can be found in \cite{KoPr09,ArBr10,NV10,KV09,NS11}.

\section{Conclusions}
\label{sec:concl}
The steep rise of the pion-photon TFF, observed by the BaBar
Collaboration, indicates the existence of an enhancement mechanism
that cannot be explained within the standard QCD scheme based on
collinear factorization.
This scheme predicts that at large values of the momentum transfer
the pion does not rebound as a \textit{unit}
but reveals its partonic structure in such a way
that the scaled TFF approaches
a constant.
Several theoretical approaches within QCD,
like LCSRs \cite{BMPS11},
Schwinger--Dyson equations \cite{RRBGGT10},
etc., are in conflict
with the dichotomous behavior of the BaBar data for the
TFFs of the pseudoscalar $\pi^0, \eta$, and $\eta'$ mesons.
Should the anomalous behavior of the $\pi^0$ BaBar data
be confirmed by independent measurements, e.g.,
by the Belle experiment,
then the BaBar effect will amount to crossing the Rubicon
and the enigma will become a real challenge
for QCD based
on collinear factorization.
Moreover, one would not be able to take recourse
for an explanation to holographic models
based on the AdS/QCD correspondence \cite{BCT11,BCT11b,GR08},
because these are incompatible
with the $\pi^0$ BaBar data at high $Q^2$.

\section{Acknowledgments}
A.\,P.\,B. acknowledges the support from the Organizing Committee
of the ``Phi-to-Psi'' Conference.
A.\,V.\,P. wishes to thank for support the Ministry of Education and
Science of the Russian Federation
(
 projects No.\ 2.2.1.1/12360 and No.\ 2.1.1/10683).
This work was supported in part by the Heisenberg--Landau Program under
Grant 2011, the Russian Foundation for Fundamental Research
(Grants No.\ 09-02-01149, 11-01-00182,  and 12-02-00613), and the
BRFBR--JINR Cooperation Program under contract No.\ F10D-002.




\end{document}